\documentclass[prc,nofootinbib]{revtex4}

\usepackage{epsfig,amsmath}

\graphicspath{{pics/}}    % look pictures from these subdirectories

\def\dndy{\frac{\mathrm{d}N_i}{\mathrm{d}y}}

\def\NS{N_\mathrm{S}}
\def\NP{N_\mathrm{P}}
\def\NPC{N_\mathrm{PC}}
\def\ssbar{{\rm s}\bar{\rm s}}

\def\d{{\rm d}}

\def\lla{\left\langle}
\def\rra{\right\rangle}
\def\gs{\gamma_S}
\def\agev{$A$ GeV}

\def\fm3{$\mathrm{fm}^3$}
\def\f3{\mathrm{fm}^3}

\def\be{\begin{equation}}
\def\ee{\end{equation}}

\begin{document}
%*****************************************************************

\title{Centrality dependence of strangeness production in
  heavy-ion collisions as a geometrical effect of core-corona superposition} 

\author{F. Becattini}\affiliation{Universit\`a di 
 Firenze and INFN Sezione di Firenze, Florence, Italy} 
\author{J. Manninen}\affiliation{INFN Sezione di Firenze, Florence, Italy}

\begin{abstract}
It is shown that data on strange particle production as a function of centrality 
in Au-Au collisions at $\sqrt s_{NN}= 200$ GeV can be explained with a 
superposition of emission from a hadron gas at full chemical equilibrium (core) and 
from nucleon-nucleon collisions at the boundary (corona) of the overlapping region of 
the two colliding nuclei. This model nicely accounts for the enhancement of $\phi$
meson and strange particle production as a function of centrality observed in 
relativistic heavy ion collisions at that energy. The enhancement is mainly a 
geometrical effect, that is the increasing weight of the core with respect to 
corona for higher centrality, while strangeness canonical suppression in the core 
seems to play a role only in very peripheral collisions. This model, if confirmed 
at lower energy, would settle the long-standing problem of strangeness 
under-saturation in relativistic heavy ion collisions, parametrized by $\gs$. 
Furthermore, it would give a unique tool to locate the onset of deconfinement in 
nuclear collisions both as a function of energy and centrality if this is to be 
associated to the onset of the formation of a fully equilibrated core. 
\end{abstract}

\maketitle

%*********************************************************************
\section{Introduction}
\label{intro}
%*********************************************************************

One of the observed features of hadron production in relativistic heavy ion 
collisions 
is the deviation from full chemical equilibrium of particles containing
strange constituent quarks. This is described by a phenomenological factor $\gs^{n_s}$
($n_s$ being the number of strange quarks in the given hadron species) which 
multiplies the equilibrium abundance of hadrons and turns out to be generally
$<1$. Actually, $\gs$ shows a mild increase as a function of centre-of-mass energy
for central collisions from 0.65 at $\sqrt s_{NN} \sim 4.5$ to about 1 at 
200 GeV \cite{bm}, according to most analyses \cite{various}. However, even at 
the largest energy of Au-Au collisions, $\gs$ turns out to be less than 1 in 
peripheral collisions, showing a monotonically decreasing trend going from central
to peripheral \cite{star,cley,bm}. 
 
The idea that $\gs$ could be the effect of superposing particle production from 
different sources in a single heavy ion collision was put forward in refs.~ 
\cite{becahi3,hohne}. 
In ref.~\cite{becahi3,becahi4}, the multiplicities of various hadron species in 
Pb-Pb collisions at SPS were described well with this core-corona picture assuming 
that corona is a halo of single nucleon-nucleon collisions where 
produced particles escape the interaction region unscathed, while the core gives 
rise to a completely equilibrated hadron gas, i.e. with $\gs=1$. Since strangeness 
production is suppressed in NN collisions with respect to a fully equilibrated hadron 
gas, while temperature is almost the same \cite{becahh,becahi4}), if such single 
NN collisions accounts for a significant fraction of total particle production, a 
global fit to one hadron-resonance gas would actually find $\gs$ significantly 
less than 1. In ref.~\cite{hohne}, the authors 
assume that a string percolation process gives rise to a large cluster in the core 
of the nuclear overlapping region and smaller clusters in the outer region (henceforth
referred to as corona), eventually decaying into hadrons according to the statistical
model {\em ansatz}. With this core-corona superposition scheme, and assuming $\gs=1$
the authors could reproduce the centrality dependence of $K/\pi$ ratio at SPS and
RHIC because small corona clusters suffer the so-called canonical suppression effect. 

This core-corona model has been applied to other observables. It has been found to 
be able to describe rapidity densities of charged hadrons in Au-Au collisions 
\cite{bozek}. More detailed analysis found out that the rapidity densities of various 
hadron species as a function of centrality in Au-Au collisions at 200\agev at RHIC 
as well as nuclear modification factors are well described with the EPOS model in a 
core-corona scheme\cite{werner}. Finally, it has been taken into account also for 
analysis of $J/\psi$ production within the statistical hadronization model \cite{pbm}.

In this paper, we will show that this core-corona superposition nicely accounts for
strangeness enhancement as a function of centrality observed at RHIC. The key probe 
which demonstrates the viability of this picture is the $\phi$ meson, which, being 
a completely neutral particle, does not feel canonical suppression. This, as we will 
discuss in detail in Sect.~\ref{coco}, favours the picture of corona as originated from 
NN collisions rather than small clusters hadronizing into a fully equilibrated hadron 
gas. In fact, in the statistical hadronization model, sufficiently small clusters 
entail canonical suppression for open strange particles, but not for those with 
hidden strangeness.

In ref.~\cite{becahi4}, we fitted the number of single nucleon-nucleon collisions 
taking place in the corona as a free parameter. In this paper, we will assume a
definition of corona as those nucleons which undergo one collision and show that
this succesfully accounts for the $\phi$ meson centrality dependence. Indications
that strangeness suppression is related to the number of multiply colliding nucleons
were found in ref.~\cite{cley} and, very recently, in ref.~\cite{timmins}; in this
work, we clarify this relation. 

The basic ideas and conclusions discussed here have been reported earlier in 
ref.~\cite{bmqm08}; in this work, we expand, explain and update our analysis.

%**********************************************************************
\section{Statistical model and canonical suppression}\label{statm}
%**********************************************************************

Statistical model analyses \cite{baran,star,bm} in Au-Au collisions at 200\agev~with 
mid-rapidity densities find that the chemical freeze-out 
temperature as well as the baryon chemical potential are constant throughout the 
accessible centrality range implying that the thermodynamical state of the produced 
matter at mid-rapidity does not depend on centrality at chemical freeze-out. Not
so for the strangeness under-saturation parameter $\gs$ which is found to be significantly
less than 1 in peripheral collisions.

It has been argued \cite{redlich} that $\gs<1$ is an effect of so-called canonical 
suppression effect. Namely, strange particles are suppressed with respect to their 
expected yield in a grand-canonical ensemble (or thermodynamic limit) because 
strangeness is exactly vanishing within a small volume, called strangeness 
correlation volume (SCV), which does not coincide with the volume of the average
fireball at mid-rapidity or the global freeze-out volume of fireballs. 
Therefore, going from pp collisions to central heavy ion collisions through 
peripheral ones, one should observe a relative enhancement of strange particles 
due to approaching the thermodynamic limit, which is hierarchical: $\Omega$ yield 
increases faster than $\Xi$ which increases faster than $\Lambda$'s or kaons.
Yet, although this hierarchy of enhancements has been observed both at SPS \cite{na57}
and RHIC \cite{starenhanc}, in neither case the natural saturation expected when SCV 
attains a sufficiently large value is seen. In fact, this means that the SCV 
would only reach its saturation value (the one sufficient for the system to be essentially 
grand-canonical) at RHIC precisely in central collisions, where $\gs \simeq 1$.
Therefore, we think that canonical suppression is quite an unnatural explanation 
of the data, as pointed out by many \cite{bmqm08,nuxu,timmins,mohanty}.

\begin{figure}[!t]
  \epsffile{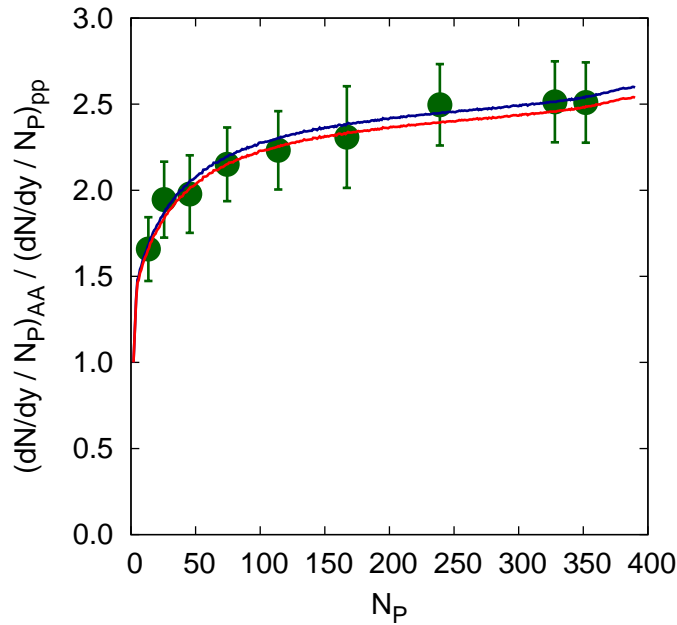}
\caption{Ratio $R_A$ (see text for definition) for the $\phi$ meson 
in Au-Au collisions at $\sqrt s_{\mathrm{NN}} = 200$ GeV. The data points are from STAR 
 \cite{starphi} while the lines are calculated according to the Eq.~(\ref{easy2}) 
 by fixing $A$ in the 2nd most central and in the mid-peripheral bin.}
\label{phi}
\end{figure}

There is, however, a clearcut probe to test the canonical suppression picture and 
this is $\phi$ meson. It is not an open strange particle, thus it is not canonically 
suppressed, yet, being a $\ssbar$ state, it must be $\gs^2$ suppressed. Furthermore, 
$\phi$ meson has almost no feeding from heavier species, so it does not suffer 
canonical suppression even indirectly as a decay product of open strange particles.
It was pointed out quite early \cite{sollfrank} that a statistical model with canonical 
suppression mechanism, i.e. with SCV as additional parameter, would have not been
able to explain the deviation of the $\phi$ meson yield from its grand-canonical 
value and this has been demonstrated in fits to NA49 multiplicities \cite{becahi3}.
Recently, STAR collaboration has measured the mid-rapidity densities of $\phi$
meson very accurately and the observed pattern as a function of centrality clearly
shows (see fig.~\ref{phi}) that these do not scale linearly with the number of
participants, rather the ratio to pp value increases rapidly at very peripheral 
collisions slowly saturating thereafter. This non-linear increase cannot be 
accounted for by a variation of the chemical freeze-out temperature because this is 
constant as a function of centrality, as has been mentioned. The only way to 
accommodate the $\phi$ meson behaviour in a statistical model fit is to introduce a 
$\gs$ factor, which is found to be significantly less than 1 in peripheral collisions.

Instead of introducing an {\em ad hoc} parameter to describe centrality dependence
of $\phi$ meson and other strange particle yields, we can try to explain strange 
phase space under-saturation as an effect superposing two different particle sources, 
as has been mentioned in the Introduction: a fully equilibrated core (i.e. a hadron 
gas with $\gs=1$) and single NN collisions in the corona. The appearance of $\gs$ in global 
statistical model fits is owing to the suppression of relative strangeness production 
(with respect to hadron gas in full equilibrium) in the unavoidably present single
NN collisions in the corona. This picture would naturally account for the decrease
of fitted $\gs$ in peripheral collisions, where the core is smaller and the importance
of single NN collisions grows.  

It should be pointed out that statistical model fits to pp collisions generally 
find temperatures only 10\% higher than in heavy ion collisions at the 
same beam energy but consistently lower $\gamma_{S\,{\rm pp}} \simeq 0.5$, i.e. about 
a factor 2 smaller \cite{becahi4}. This explains why, in this core-corona model, 
no effect is seen in global fits on centrality dependence of temperature but a 
significant dependence of $\gs$.

Finally, the shape of normalized $\phi$ mid-rapidity density in fig.~\ref{phi} also 
tells us that the corona cannot be really made of small clusters hadronizing into
a fully equilibrated hadron gas at the same temperature of the core. Indeed, in 
this case, $\gs=1$ and one could account only for the suppression of open strange 
hadrons but not of $\phi$.

%**********************************************************************
\section{Core-corona superposition}\label{coco}
%**********************************************************************

In the following, we will introduce a very simple model which is suitable to study 
core-corona superposition in heavy ion collisions and, particularly, to estimate the 
relative weight, in terms of particle production, of core and corona as a function 
of centrality. 

It has been observed\cite{Elias:1979cp,Back:2004mr} that the number of charged hadrons 
emitted in hadron-nucleus (hA) collisions, compared with pp collisions at the same 
beam energy, scales as:
\begin{equation}\label{pA}
 \lla{\frac{\d N^{ch}}{\d y}}\rra_{\mathrm{hA}}/ \lla{\frac{\d N^{ch}}{\d y}}\rra_{\mathrm{pp}} 
 = \frac{\NP^{\mathrm{hA}}}{\NP^{\mathrm{pp}}} = \frac{\NP}{2} = \frac{N_{\mathrm{coll}}+1}{2},
\end{equation}
where $N_{\mathrm{coll}}$ is the number of collisions. Based on the above formula, the 
rapidity density of a given hadron species $i$ in heavy ion collision at mid-rapidity 
is written as a sum of two contributions:
\begin{eqnarray}\label{sp}
\lla{\dndy}\rra &=& \frac{\NPC}{2} \lla{\dndy}\rra_\mathrm{NN} + 
 \lla{\dndy}\rra_\mathrm{core} \nonumber\\
&\simeq& \frac{\NPC}{2} 
 \lla{\dndy}\rra_\mathrm{pp} + \lla{\frac{\d N_i}{\d y}}\rra_\mathrm{core}.
\end{eqnarray}
In the above equation, $\NPC$ is the mean number of participants in the low 
density corona, which undergo single nucleon-nucleon collisions whose 
produced particles escape the interaction region unscathed. Since at high energy 
hadron production in neutron-neutron as well as in neutron-proton collisions closely 
resembles that in proton-proton collisions at the same beam energy, the second 
(approximate) equality in Eq.~(\ref{sp}) holds true. The rapidity density in pp 
collisions is measured, while that in the core is assumed to be that of a completely
equilibrated hadronic source.

At very high energy (typically RHIC energies) where particle rapidity distribution is 
wide, the second term on the right hand side of Eq.~(\ref{sp}) can be written 
\cite{becahi4}:
\begin{equation}\label{core}
 \lla{\frac{\d N_i}{\d y}}\rra_\mathrm{core} = V \rho_0 
 \lla{\frac{\d n_i}{\d y}}\rra_\mathrm{core}
\end{equation}
where $n_i$ is the density of hadron species $i$ in the fully equilibrated average
fireball of freeze-out volume $V$ and $\rho_0$ is the density of fireballs per unit 
of rapidity at mid-rapidity. For a collision of nuclei with mass numbers $A$ and $B$, 
the volume of the core at freeze-out can be estimated:
\begin{equation}\label{vola}
 V = f (V_0 - \delta V_0) 
 \approx f \left( \frac{\NP^A+\NP^B}{2n_0} - \frac{\NPC^{A}+\NPC^{B}}{2n_0} \right)
 = \frac{f}{2n_0}(\NP-\NPC).
\end{equation}
In Eq.~(\ref{vola}) $\NP$ is the number of participants, $\NPC$ is the number of 
participants in the corona; $V_0\approx \NP / 2n_0$ ($n_0$ being some initial 
density related to nuclear density) is the average of the initial overlap volume of 
the two colliding nuclei; $\delta V_0 \approx \NPC / 2n_0$ is the average volume of 
the corona, $f$ is a ``growth factor" which takes into account the expansion of the 
system between the initial overlap time and freeze-out. It should be stressed that
the essential point here is the proportionality between the core volume and the 
number of participants of the core obtained as difference between the total number
and that of the corona; indeed, as we will see, the knowledge of the parameters $f$ and
$n_0$ is not needed for our purpose. \\
Plugging Eq.~(\ref{vola}) into (\ref{core}) we obtain:
\begin{equation}\label{core2}
 \lla{\frac{\d N_i}{\d y}}\rra_\mathrm{core} = \frac{f \rho_0}{2n_0}(\NP-\NPC) 
 \lla{\frac{\d n_i}{\d y}}\rra_\mathrm{core}.
\end{equation}
Since charged particle multiplicity scales linearly with the number of participants 
in heavy ion collisions, we are led to conclude that the factor $f \rho_0/2n_0$ in 
Eq.~(\ref{core2}) is independent of $\NP$, i.e. independent of centrality, provided 
that $\lla{\d n_i / \d y} \rra_\mathrm{core}$ is in turn independent of centrality.
As we will see, this condition holds true because of the observed independence of
freeze-out parameters on centrality.

Putting Eq.~(\ref{core2}) into (\ref{sp}) and dividing both sides by 
$\NP \lla {\d n /\d y}\rra_\mathrm{pp}/2$, we obtain:
\be\label{easy}
 R_A = \frac{2\lla{\dndy}\rra_\mathrm{AA}}{\NP \lla{\dndy}\rra_\mathrm{pp}}
 = \frac{2f \rho_0}{2n_0}\frac{\lla{{\frac{\d n_i}{\d y}}}\rra_\mathrm{core}}
 {\lla{\dndy}\rra_\mathrm{pp}}\left( 1- \frac{\NPC}{\NP} \right) + \frac{\NPC}{\NP}
 = \frac{\NPC}{\NP} + A \left( 1-\frac{\NPC}{\NP}\right). 
\ee
The factor $A$ embodies all unknown parameters and depends linearly on particle 
density in the core. If this factor was independent of $\NP$, the whole centrality
dependence of the $R_A$ ratio would be given by quantity related to the geometry
of the collision. Indeed, we now need to define the NN collisions which form 
the corona, or, in other words to define $\NPC$.

Based on Eq.~(\ref{pA}) for hA collisions, where the impinging and target nucleons 
collide only once, we choose to define $\NPC$ as the number of nucleons colliding 
only once. Explicitly: 
\begin{equation}\label{ns}
 \NPC = \NS^A + \NS^B
\end{equation}
in which $\NS^{A,B}$ are the numbers of such singly colliding nucleons from nuclei 
$A$ and $B$ respectively \footnote{In ref.~\cite{bmqm08} we have used a 
different definition of $\NPC$, i.e. twice the minimum between $\NS^A$ 
and $\NS^B$. We find that the present definition is conceptually more 
satisfactory and better motivated experimentally. However, they eventually 
lead to very similar results.}.

Armed with the definition (\ref{ns}) of single collisions in the corona, and 
with the equation (\ref{easy}), we are now in a position to test the model on the 
data by calculating $\NP$ and $\NPC$ with the Glauber model.

%-----------------------------------------------------------------------------------
%------------------------------------------------------------------------------------

%**********************************************************************
\section{Data analysis and results}
%**********************************************************************

We have emphasized that the $\phi$ meson provides us with an excellent probe to test 
models of strangeness suppression, and so does it for the core-corona model and the 
Eq.~(\ref{easy}), that we rewrite here:
\begin{equation}\label{easy2}
 R_A = \frac{2\lla{\dndy}\rra_\mathrm{AA}}{\NP \lla{\dndy}\rra_\mathrm{pp}}
 = A + \frac{\NPC}{\NP}(1-A). 
\end{equation}
Since $\phi$ is immune from finite size effects, i.e. canonical suppression, the factor 
$A$ is independent of centrality because chemical freeze-out temperature is found to be 
centrality-independent. We can then fix the factor $A$ from the data in one centrality 
bin and see how well the normalized yield $R_A$ is described at other centralities by
calculating $\NP$ and $\NPC$ with a Glauber Monte-Carlo simulation. 

\begin{figure}[!htb]
    \epsffile{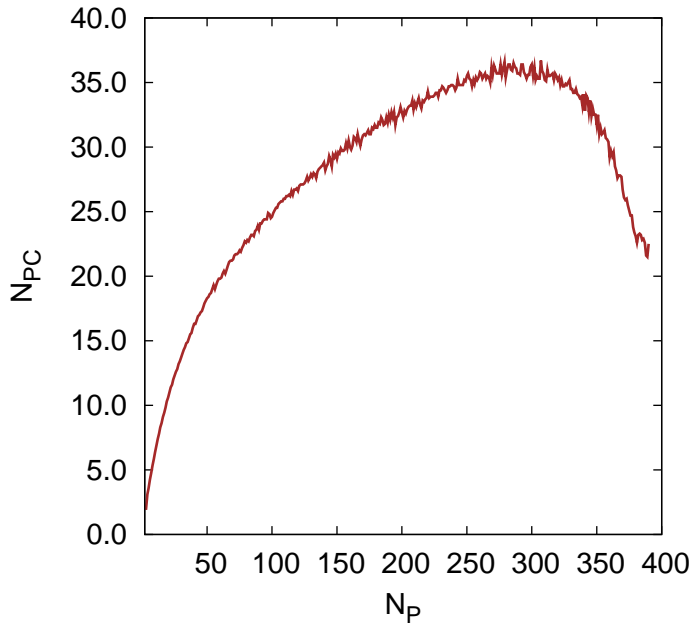} 
\caption{Number of nucleons $\NPC$ undergoing one collisions defining the corona
as a function of $\NP$ in Au-Au collisions at $\sqrt s_{NN} = 200$ GeV.}
\label{npc}
\end{figure}

For our Glauber Monte-Carlo calculation we have implemented essentially the same 
algorithm used in ref.~\cite{steinberg}. Our calculated number of participants in the
corona according to formula (\ref{ns}) is shown in Fig.~\ref{npc}. In Fig.~\ref{phi}
$R_A$ for the $\phi$ meson measured at RHIC at $\sqrt s_{NN}=200$ GeV \cite{starphi} 
is compared with the theoretical calculation. The two lines are calculated according 
to the Eq.~(\ref{easy2}) by fixing the factor $A$ from two different centrality bins. 
We can see that in both cases the agreement between model and experiment is excellent. 

%-----------------------------------------------------------------------------------
%------------------------------------------------------------------------------------

%-----------------------------------------------------------------------------------
\begin{figure}[htb]
$\begin{array}{c@{\hspace{-0.05cm}}c}
  \epsffile{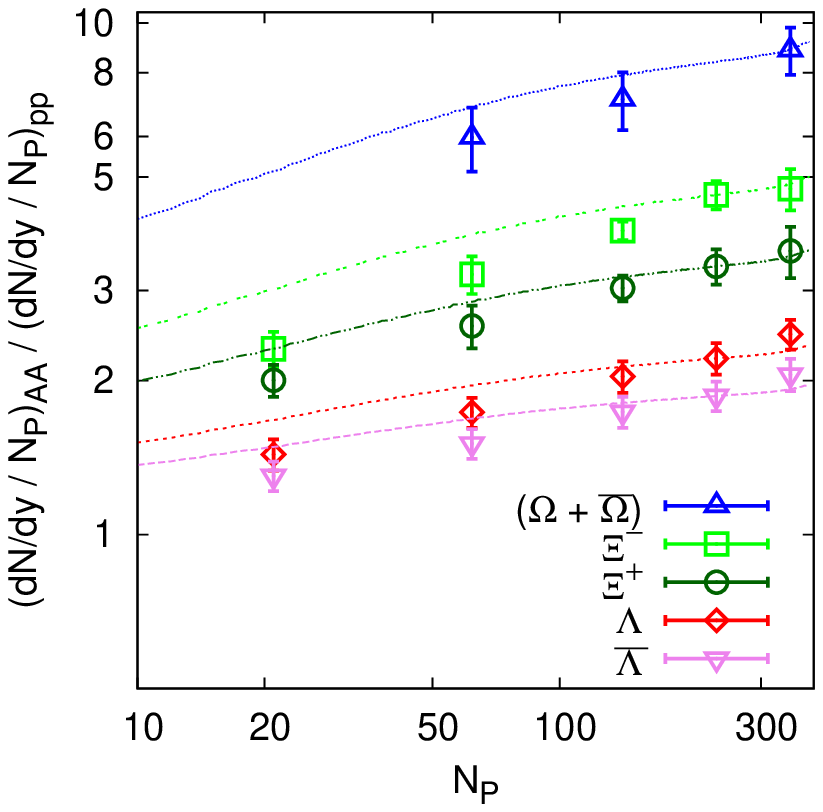} & \epsffile{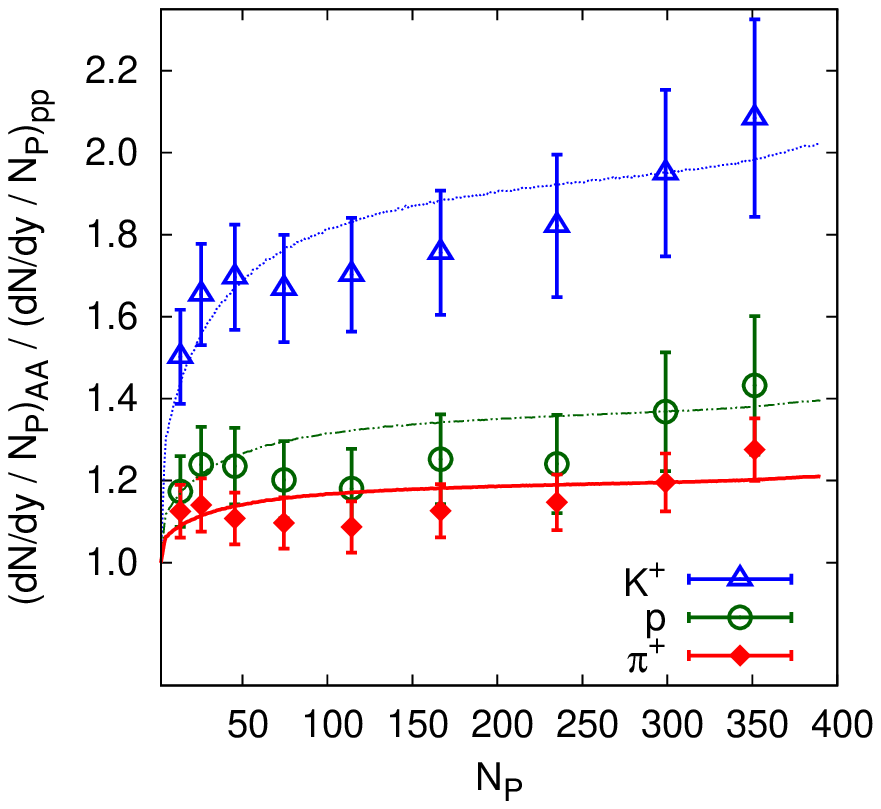} \\
\end{array}$
\caption{{\bf LEFT}: Ratio $R_A$ (see text for definition) for hyperons measured
in Au-Au collisions at $\sqrt s_{\mathrm{NN}} = 200$ GeV. The data points are from STAR 
 \cite{starenhanc} while the lines are calculated according to the Eq.~(\ref{easy2}) 
 by fixing $A$ in the most central bin.\\ 
  {\bf RIGHT}: The same quantity $R_A$ for $\pi^+$, $K^+$ and $p$ in Au-Au collisions at 
  $\sqrt s_{\mathrm{NN}} = 200$ GeV. The data points are from STAR \cite{stable} while the 
  lines are calculated according to the Eq.~(\ref{easy2}) by fixing $A$ to the 2nd most 
  central bin.}
\label{hyp}
\end{figure}
%------------------------------------------------------------------------------------

On the left panel of Fig.~\ref{hyp}, similar curves are shown for $\Lambda$, 
$\bar{\Lambda}$, $\Xi^{\pm}$ and $\Omega+\bar{\Omega}$ compared with $R_A$'s measured
at RHIC \cite{starenhanc} at the same beam energy. The factor $A$ is fixed to the
2nd most central bin\footnote{except $\Omega+\bar{\Omega}$ to the most central bin} 
and, as expected, the model overshoots the data in the most peripheral 
collisions. This is because the factor $A$ is not in fact independent of centrality 
and should decrease in very peripheral bins due to strangeness canonical suppression. 
We note in passing that the difference between the curve and the points gives then 
quantitative information about the volume of the average fireball at mid-rapidity.

\begin{figure}[!t]
    \epsffile{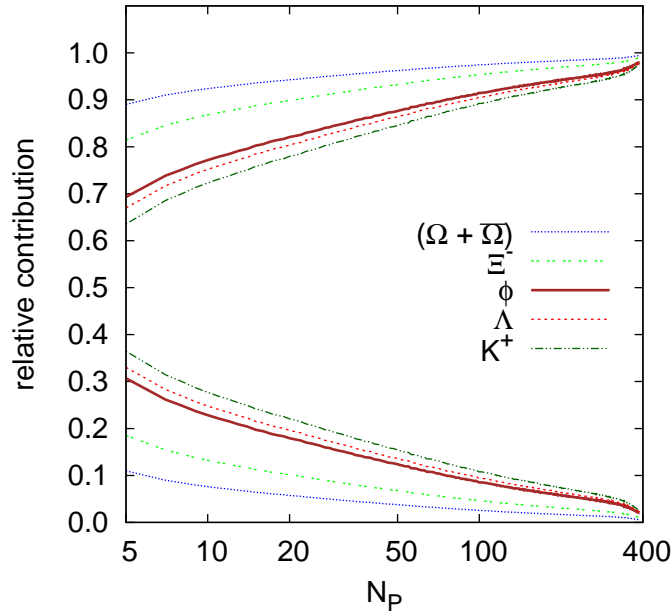} 
\caption{Fraction of produced particles coming from core (upper lines) and corona 
 (lower lines) as a function of centrality.}
\label{weight}
\end{figure}

In Fig.~\ref{weight} we show the fraction of particle production from core and corona 
as a function of centrality for different particle species. This plot shows that 
the hierarchy of enhancement slopes observed at RHIC is nicely reproduced 
in this approach. According to Eq.~(\ref{easy2}), the higher is $A$, the steeper is
the increase of $R_A$ as a function of $\NP$. Since $A$ depends on particle species 
through the ratio:
$$
\frac{\lla{{\frac{\d n_i}{\d y}}}\rra_\mathrm{core}}
 {\lla{\dndy}\rra_\mathrm{pp}}
$$
the observed hierarchy simply reflects the hierarchy of ratios of mid-rapidity 
densities in heavy ion to NN collisions, therefore $\Omega > \Xi^- > \bar \Xi^+ >
\phi > \Lambda > \bar \Lambda$ as it results from Fig.~\ref{weight}. 

%-----------------------------------------------------------------------------------
%------------------------------------------------------------------------------------

\begin{figure}[!t]
$\begin{array}{c@{\hspace{-0.05cm}}c}
       \epsffile{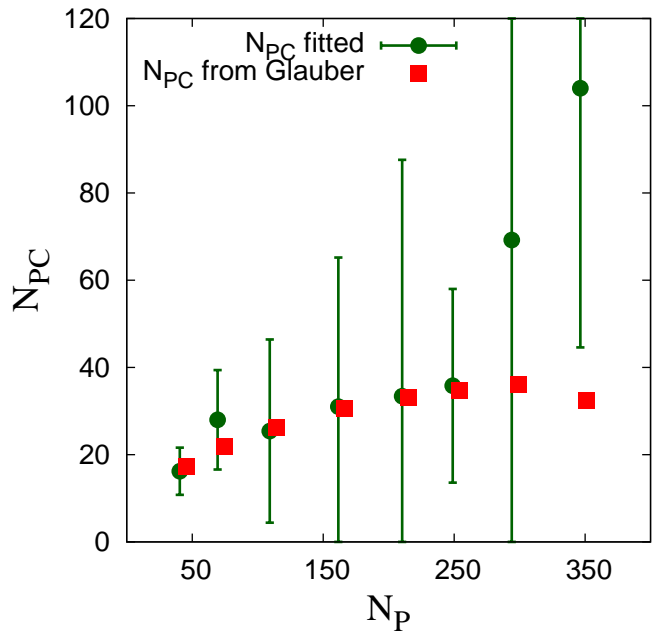} \\
\end{array}$
\caption{Number of corona participants $\NPC$ at different centralities. The square 
 dots denote the values calculated from Glauber model while the round ones 
 denote the values arising from fitting $\NPC$ as a free parameter.
The round symbols are shifted 5 units of $\NP$ rightward for clarity.}
\label{fit}
\end{figure}

The model can be further tested with the RHIC data directly with the Eq.~(\ref{sp})
in the statistical hadronization model framework (see ref.~\cite{bm}).  
The hadron radiation from the corona part can be estimated by taking the experimental 
rapidity densities of different hadron species $i$ in pp 
collisions~\cite{Adams:2003xp,starphi,Abelev:2006cs} measured by the STAR 
collaboration and multiplying these with the number of corona participants $\NPC$. 
This number can be determined in a twofold way: by fitting it as a free parameter 
or calculating it directly from the Glauber model as the total number of singly
colliding nucleons as in Eq.~(\ref{ns}). In the first case we will have 4 free 
parameters that must be fitted to the measured rapidity densities at different 
centralities while in the second case we have 3 free parameters only. The fitted and 
calculated $\NPC$ as a function of $\NP$ at different centralities are shown in 
fig.~\ref{fit}. One can see that the fitted and calculated number of single collisions 
agree very well with each other in peripheral and semi-central collisions while in 
central collisions the relative emission from the corona is too small for us to 
reliably fit the number of single collisions. We also remark that hadron yields are 
described better in this two-component formalism than the conventional statistical 
hadronization model analysis~\cite{bm} with the $\gamma_S$ parameter. 

%-----------------------------------------------------------------------------------
%------------------------------------------------------------------------------------

%**********************************************************************
\section{Summary and conclusions}
%**********************************************************************

In summary, we confirm our early finding \cite{bmqm08} that strangeness enhancement 
from peripheral to central
relativistic heavy ion collisions at RHIC at $\sqrt s_{NN}=200$ GeV can be well 
described by a model where particle production arises from two sources: a fully
equilibrated core at a temperature of $\sim 160$ MeV as in the statistical model
and an outer region of single NN collisions, called corona. Since in NN collisions
relative strangeness production is suppressed with respect to a fully equilibrated
grand-canonical core, the observed enhancement going from peripheral to central 
stems from the increased weight of the core with respect to corona. The enhancement 
is thus mainly a geometrical effect and it is not driven by the conservation of net 
strangeness within small regions in the core itself (canonical suppression picture): 
this effect shows up only in very peripheral collisions where the whole core's volume 
is presumably small. 

The key probe for our argument is the $\phi$ meson which
is immune from canonical suppression and has essentially no feeding from high-lying
resonances. The observed rise of relative $\phi$ yield as a function of centrality 
is an unambiguous signal that the increased production of strangeness from peripheral 
to central collisions is not an effect of strangeness conservation nor can it be 
explained by an increase of temperature, which is found to be constant throughout. 
The $\phi$ enhancement also favors the idea that corona consists of independent 
NN collisions rather than small clusters hadronizing in full chemical equilibrium. 
In the latter case, with a hadronization temperature at the same value of around 
165 MeV (as confirmed by analysis of pp collisions), $\phi$ normalized production 
would be flat as a function of centrality. 

The success of this description indicates, as pointed out, that a fully equilibrated 
core is formed at any centrality in Au-Au collisions at $\sqrt s_{NN} = 200$ GeV.
Arguably, this should occur also at lower energies and so core-corona superposition
is likely to explain strangeness enhancement observed by NA57 experiment in 
Pb-Pb collisions at $\sqrt s_{NN}=17.3$ GeV. It becomes then crucial, in order to 
locate the onset of full chemical equilibrium in the core, to study the production 
of hyperons and chiefly $\phi$ mesons as a function of centrality in the energy 
range from few GeV's to 20 GeV. When, at some low energy or centrality, the data 
will start to be well reproduced by a fully equilibrated core, that might be be 
the point where deconfinement has occurred.

Finally, the superposition of core and corona appears to be a general feature
of relativistic heavy ion collisions which should be taken into account for the
analysis of all observables besides particle chemistry. This has been pointed out
in refs.~\cite{bozek,werner} and in a very recent analysis \cite{bozek2}. The 
definition of the corona is not unique and our results indicate that it should
be based on NN collisions rather than local density \cite{bozek,werner,pbm}. 
In this regard, a global fit to particle spectra \cite{bozek2} also favors 
our definition of corona as being made of those nucleons undergoing one collision.

%**************************************************************************
\section*{Acknowledgments}
%**************************************************************************

We are greatly indebted with H. Caines and J. H. Chen from STAR collaboration
for their help in handling particle production measured by STAR and useful
discussions. We acknowledge stimulating and very useful discussion with P. 
Steinberg regarding Glauber model calculations.

%************************************************************************

\end{document}